\documentclass[aps,prl,twocolumn]{revtex4-1}
\usepackage{graphicx}

\begin{document}
\title{Isotope Effect on Electron-Phonon Interaction in Multiband Superconductor MgB$_2$}
\author{Daixiang Mou, Soham Manni, Valentin Taufour, Yun Wu, Lunan Huang, S. L. Bud'ko, P. C. Canfield, Adam Kaminski}
\affiliation{
\\$^{1}$Division of Materials Science and Engineering, Ames Laboratory, Ames, Iowa 50011, USA
\\$^{2}$Department of Physics and Astronomy, Iowa State University, Ames, Iowa 50011, USA
}

\begin{abstract}
We investigate the effect of isotope substitution on the electron-phonon interaction in the multi-band superconductor MgB$_2$ using tunable laser based Angle Resolved Photoemission Spectroscopy. The kink structure around 70 meV in the $\sigma$ band, which is caused by electron coupling to the $E_{2g}$ phonon mode, is shifted to higher binding energy by $\sim$3.5 meV in  Mg$^{10}$B$_2$ and the shift is not affected by superconducting transition.
These results serve as the benchmark for investigations of isotope effects in known, unconventional superconductors and newly discovered superconductors where the origin of pairing is unknown.
\end{abstract}
\maketitle

The discovery of the isotope effect in conventional superconductor played a key role in establishing Bardeen-Cooper-Schrieffer(BCS) theory \cite{Maxwell1950,Reynolds1950,Frohlich1950,Bardeen1957}, in which electrons form Cooper pairs by coupling via lattice vibrations (phonons). The isotope effect is regarded as a decisive method to determine whether or not the pairing in newly discovered superconductors is phonon mediated, or of unconventional (e. g. electron-electron) origin. This approach worked beautifully in case of the MgB$_2$ superconductor \cite{Nagamatsu2001,Budko2001} and immediately established electron-phonon coupling as the origin of pairing and superconductivity in this material \cite{Budko2001}. The phonon mediated superconductivity in this material was explained by multiband BCS/Eliashberg theory shortly after its discovery \cite{Budko2001,Hinks2001,Kortus2001,Choi2002}. Similar measurements have been made on cuprate high temperature superconductors, but no unambiguous conclusion has been reached because of the complexity of isotope response and inconsistent results \cite{Franck1994,Keller2005,Liu2009,Shirage2009}. Thus, MgB$_2$ is unique in its combination of very high T$_c$ values and unambiguous electron-phonon coupling.

The electron-boson interaction renormalizes the binding energy of the electrons near chemical potential, causing a ``kink" structure in the band dispersion which is a signature of a sudden change of self-energy due to coupling to the bosonic mode. Angle Resolved Photoemission Spectroscopy (ARPES), a powerful tool that can measure the band dispersion and self-energy with momentum resolution, provides a direct way to study the electron-boson interaction. The kink structure in cuprate high temperature superconductor has been intensively investigated by ARPES for years because it might contain essential information to establish the high temperature superconducting mechanism \cite{Damascelli2003}. But whether the observed kink is related to phonon or spin fluctuation is still an open question. There were several ARPES-based studies of the isotope effect in cuprates superconductors  carried out in the past \cite{Gweon2004,Douglas2007,Iwasawa2007,Iwasawa2008}. The results included observation of a very large change in band dispersion at high binding energies, interpreted as a signature of polaronic effects \cite{Gweon2004} and observation of very small effect at low energies along the nodal direction \cite{Iwasawa2008}. Due to difference in conclusions the significance and relation to high temperature superconductivity remains unclear. Surprisingly, to the best of our knowledge no direct experimental study of the isotope effect on the kink structure in a classical BCS superconductor has been reported . MgB$_2$ offers a unique opportunity due to it relatively high T$_c$ and its quasi two dimensional character of the $\sigma$ bands.

\begin{figure*}[htbp]
\centering
\includegraphics[width=0.9\textwidth]{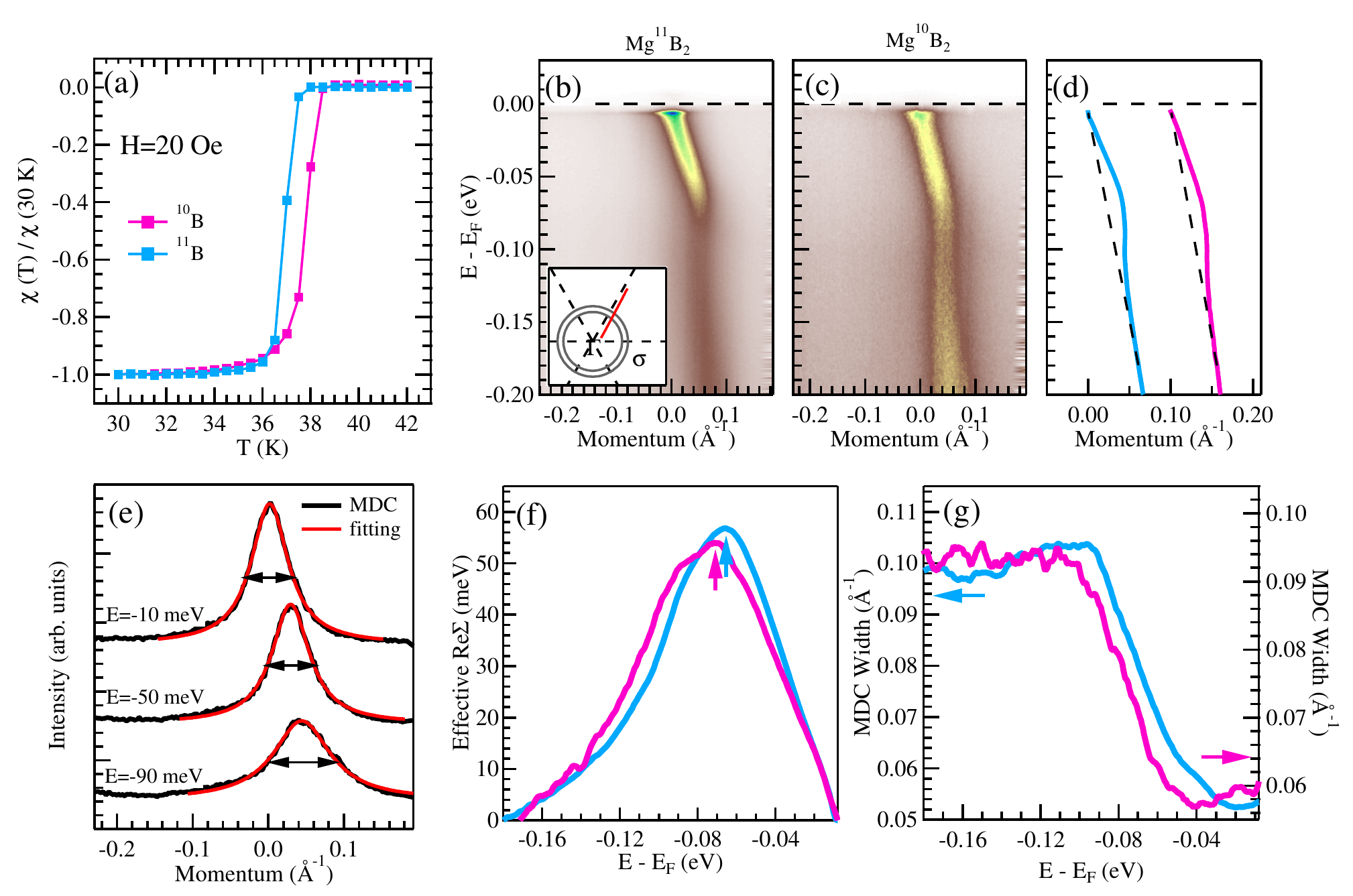}
\caption{(Color online) Isotope effect on kink structure in MgB$_2$. (a) Temperature dependence of magnetic susceptibility of Mg$^{11}$B$_2$ and Mg$^{10}$B$_2$ single crystals used in ARPES measurements. (b)-(c)  ARPES intensity plots for Mg$^{11}$B$_2$ and Mg$^{10}$B$_2$ samples along a cut that crosses $\sigma$ band measured at 15~K.  (d) Band dispersion extracted by fitting the MDC data with Lorentzian function, as shown in panel (e). The curves are shifted horizontally for clarity. Dashed lines are corresponding bare bands used to extract the Re$\Sigma$. (e) MDCs and Lorentzian fits at three sample binding energies. Vertical arrows mark the MDC width that is proportional to Im$\Sigma$. (f) Effective Re$\Sigma$ obtained by subtracting the ``bare" band dispersion (dashed lines in (d)) from extracted dispersion. Kink positions are marked with arrows. (g) Energy dependence of the MDC width.}
\end{figure*}

MgB$_2$ is a layered multiband superconductor with a transition temperature as high as 40 K \cite{Nagamatsu2001,Budko2001}. The strong electron-phonon interaction and discovery of large isotope effect on T$_c$ \cite{Budko2001,Hinks2001} make this material an ideal system for the investigation of this important topic and establish benchmarks for gaining insights to novel superconductors. Early ARPES studies of MgB$_2$ did not reveal the kink structure because of experiment resolution and sample quality and size limitations \cite{Uchiyama2002,Souma2003,Tsuda2003}. Recently, the kink structure was identified using our tunable laser ARPES system with much improved instrumental resolution \cite{Mou2015a,Mou2015b}. This provides us the opportunity to study the isotope effect on the kink structure in a prototypical conventional superconductor. In this letter, we report an ARPES study of the electron-phonon interaction in Mg$^{10}$B$_2$ and Mg$^{11}$B$_2$ samples. We will show the clear isotope shift of the kink structure on the $\sigma$ band and that this isotope shift is not significantly affected by the superconducting transition.

\begin{figure*}[htbp]
\centering
\includegraphics[width=0.9\textwidth]{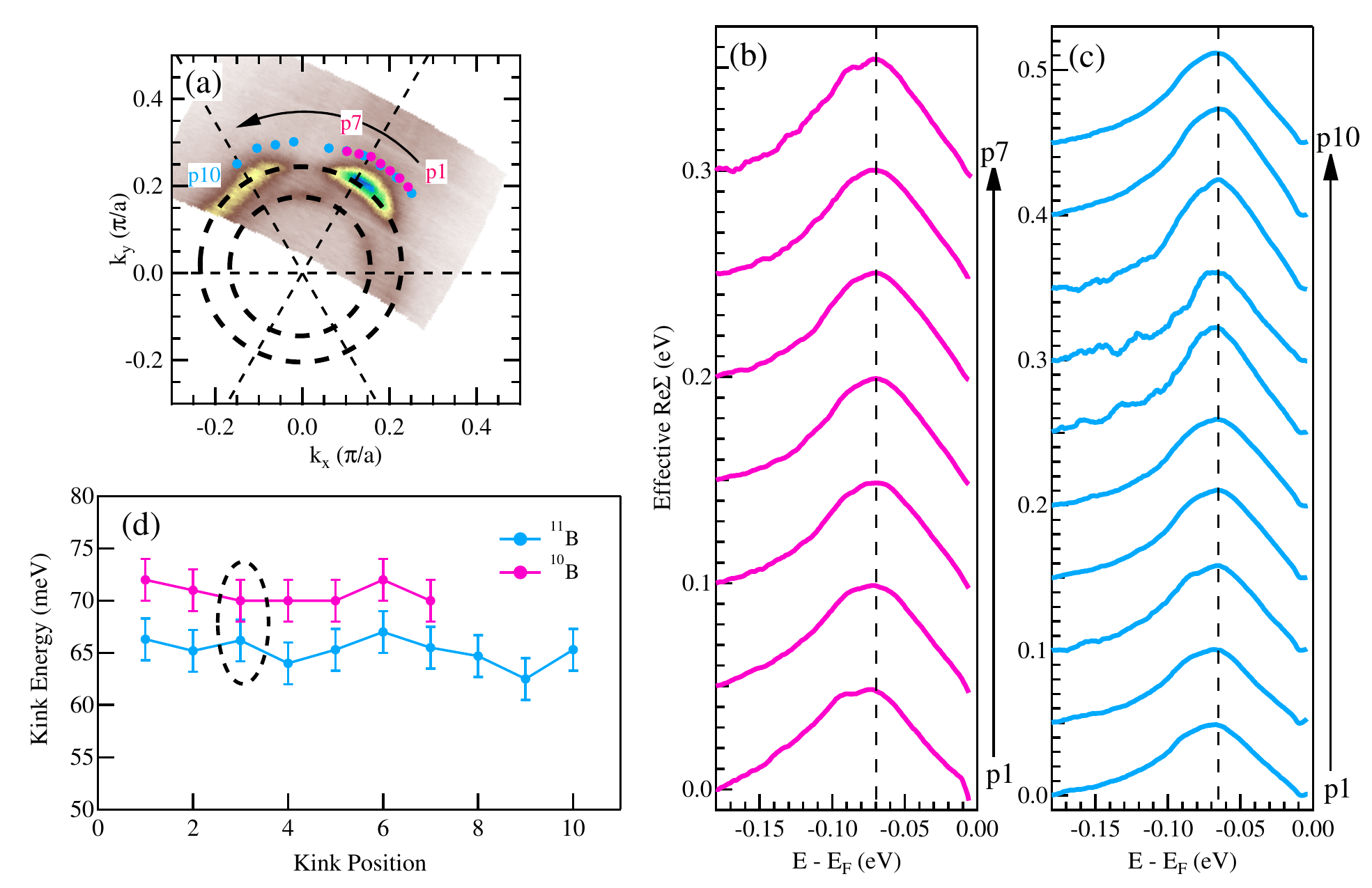}
\caption{(Color online) Momentum dependence of the kink structure. (a) Measured Fermi surface of Mg$^{10}$B$_2$ at 40 K. (b)-(c) Momentum dependence of the effective Re$\Sigma$ of outer $\sigma$ band in Mg$^{10}$B$_2$ (b) and Mg$^{11}$B$_2$ (c). Data were taken at 20 K. (d) Momentum dependence of the kink energy obtained from (b) and (c). Corresponding kink structure positions in Brillouin Zone are mark with circles in (a). The data points in Fig. 1 are circled out.}
\end{figure*}

MgB$_2$ single crystals are usually grown from a solution with excess Mg under high pressure and high temperature in BN crucibles \cite{Karpinski2007}. Unfortunately, for isotopic substitution, there are clear indications that BN is partly reacting with Mg during the high pressure/high temperature growth process.  The incorporation of BN into the high temperature melt is suggested by the growth of BN single crystals together with MgB$_2$ and is clearly implicated by the fact that MgB$_2$ crystals grown with nominally, isotopically pure $^{11}$B contain detectable and enhanced levels of $^{10}$B that appears to be coming from the BN crucibles made from natural abundance B \cite{Cubitt2003}. To avoid any contamination from natural B contained in the BN crucible, we performed the MgB$_2$ growths using MgO crucibles. Under these conditions, only single crystals of MgB$_2$ form and the isotopic purity of B can be preserved. The success of these growths also demonstrates that MgB$_2$ single crystals can be grown from a binary melt when adequate pressure is used to both control vapor pressure and shift the phase lines in the binary phase diagram.  Single crystals of both Mg$^{11}$B$_2$ and Mg$^{10}$B$_2$ were grown in the following manner using high purity Mg as well as 99.5\% isotopically pure $^{11}$B and $^{10}$B from Eagle Pitcher.  Elemental Mg and B were placed in a thin walled MgO crucible in the ratio of Mg:B 1:0.7.  Using a cubic anvil furnace, we apply around 3.5 GPa pressure on the sample at room temperature, then the temperature is increased to 1450 $^{\circ}$C over approximately 2 hours, after dwelling at 1450 $^{\circ}$C for 2 hours, the temperature is decreased to 650 $^{\circ}$C over 6 hours.  At this point the heater is turned off, quenching the growth to room temperature and pressure is decreased back to ambient. The MgO crucible is then removed from the pressure media and the excess Mg is distilled off of the MgB$_2$ crystals. 

The typical size of the samples used in ARPES measurements was $\sim$$0.3\times0.3\times0.2$ mm$^3$. Samples were cleaved \emph{in situ} at a base pressure of lower than 8 $\times$ 10$^{-11}$ Torr.
  ARPES measurements were carried out using a laboratory-based system consisting of a Scienta R8000 electron analyzer and tunable VUV laser light source \cite{Jiang2014}. The energy resolution of the analyzer was set at 1 meV and angular resolution was 0.13$^\circ$ and $\sim$ 0.5$^\circ$ along and perpendicular to the direction of the analyzer slits, respectively. All data were acquired using a photon energy of 6.7 eV; only the electronic structure of the two $\sigma$ bands around Brillouin zone center can be measured with this photon energy. Fortunately it is the $\sigma$ band that couples strongly with the phonons. The laser beam was focused to a spot with a diameter $\sim$ 30 $\mu$m. Samples were cooled using a closed cycle He-refrigerator and  the sample temperature was measured using a silicon-diode sensor mounted on the sample holder. The energy corresponding to the chemical potential was determined from the Fermi edge of a polycrystalline Au reference in electrical contact with the sample. In every temperature dependent measurement, aging effects were checked by thermal cycling. The consistency of the data was confirmed by measuring several samples.

\begin{figure}[htbp]
\centering
\includegraphics[width=0.8\columnwidth]{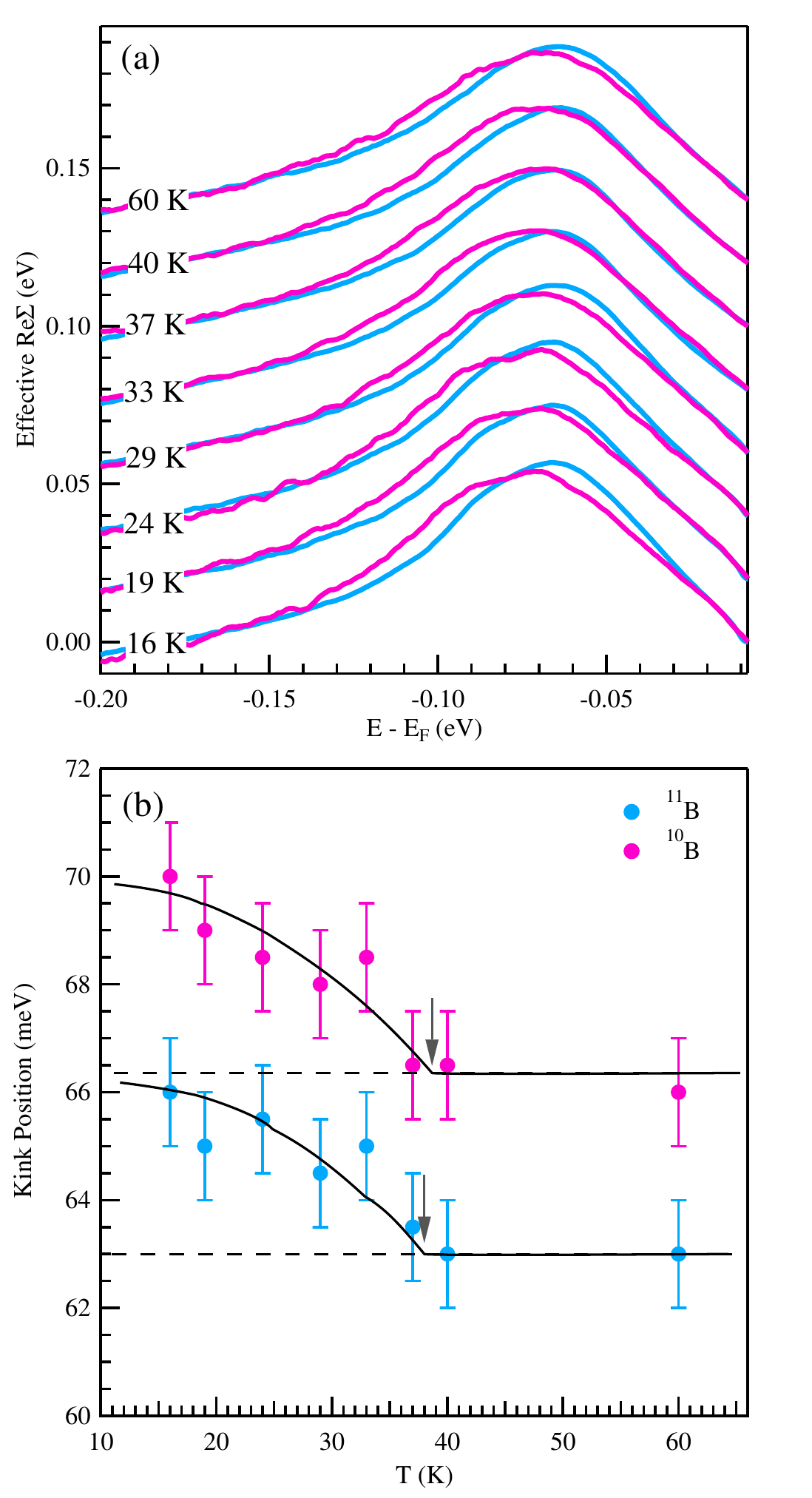}
\caption{Temperature dependence of the Re$\Sigma$ for the samples made using two different boron isotopes. (a) Effective Re$\Sigma$ for the outer $\sigma$ band extracted from data measured at different temperatures. Cut position is the same as in Fig.1. (b)  Temperature dependence of the kink energy for the two isotopes. Solid lines are guide for the eye. Arrows show corresponding transition temperatures from Fig. 1. }
\end{figure}

Fig. 1a shows the low field, temperature dependence of the magnetization for representative crystals of Mg$^{10}$B$_2$ and Mg$^{11}$B$_2$.  A clear, approximately 1.0 K isotope shift is seen in these data. It should be noted that each transition is lower by approximately 1.0 K (T$_c$($^{11}$B)=37.5 K, T$_c$($^{10}$B)=38.5 K) from what was previously reported in ref. \cite{Budko2001}.  This is due to the fact that the high pressure growth requires the use of a graphite heater in close proximity to the Mg-B melt.  This leads to a small carbon substitution ($\sim$1\%) that mildly suppresses T$_c$. The measured ARPES data in the superconducting state (T=15 K) of $^{11}$B and $^{10}$B samples are shown in Fig. 1b and 1c. We chose a particular cut position in the momentum space (see the measured Fermi surface in Fig. 2a) where the matrix element totally suppresses the intensity of the inner $\sigma$ band to simplify the data analysis. The anomaly in the band dispersion (kink) can be clearly seen directly in the the ARPES intensity plots around a binding energy of 70 meV and is caused by the coupling of the electrons to the E$_{2g}$ phonon mode as discussed in our previous report \cite{Mou2015a}. In order to quantitatively analyze the details of this kink structure, we extracted the band dispersion, as shown in Fig. 1d, by fitting the Momentum Distribution Curves (MDC) with Lorentzian lineshape (Fig. 1e). The effective real part of self-energy (Re$\Sigma$) is then obtained by subtracting a linear ``bare" band dispersion (plotted as a dashed lines in panel d) from the measured dispersion. The width of the MDCs is a measure of the imaginary part of the self energy (Im$\Sigma$) and is enhanced for the binding energies higher than the energy of the boson mode plus superconducting gap as shown in Fig. 1g. Both the shape of Re$\Sigma$ (Fig. 1f) and MDC width (Fig. 1g)  show a shift toward higher binding energy in $^{10}$B sample, providing clear evidence of the strong isotope effect in this material. Theoretically, either the peak position in Re$\Sigma$ or the midpoint of dropping edge in Im$\Sigma$ can be used to extract coupling boson mode energy, but the Re$\Sigma$ is more accurately extracted from data fits (it is easier to extract peak position than its width) and is more commonly used in literature \cite{Damascelli2003}. The energy value of the Re$\Sigma$ peak is 66.5 meV for $^{11}$B sample and  70 meV for $^{10}$B sample respectively. Therefore the effects of different isotope mass on the energy of the phonon mode amounts to shift of $\sim$ 3.5 meV

In order to confirm that the observed isotope energy shift does not arise due to some small angular misalignment associated with two measurement of two separate crystals, we measured the momentum dependence of the kink structure in both samples over a wide angular range. Following the same procedure as described above, we extracted the effective Re$\Sigma$ along different cuts in momentum space in Fig. 2b ($^{10}$B sample) and Fig. 2c ($^{11}$B sample) and the momentum dependent kink energy in Fig. 2d. The energy position of kink structure is nearly momentum independent in both samples demonstrating that the reported here shift of the kink energy attributed to the isotope effect is not due to an artificial caused by slight misalignment during measurement, which was a likely culprit in early isotope experiments in cuprates \cite{Gweon2004}.

Finally we turn to the temperature dependence of the isotope effect. We compare the extracted Re$\Sigma$ of the outer $\sigma$ band at different temperatures in Fig. 3a. The shift of the peak structure is obvious for all measured temperatures  up to 60 K. In fig. 3b, we extracted temperature dependent  peak position of Re$\Sigma$ in both samples. The temperature dependence of the kink position in $^{11}$B samples was already discussed in our previous report \cite{Mou2015a}. The kink energy remains constant in the normal state and starts to shift to high binding energy gradually below T$_c$ with a total shift of $\sim$3 meV at the lowest measured temperature. The kink position in $^{10}$B sample has a similar temperature dependence as that in $^{11}$B sample and also shifts towards a higher binding energy by $\sim$ 3.5 meV, indicating the isotope effect on kink structure is not affected by the superconducting transition in MgB$_2$. The lifetime of the quasiparticles in the $\sigma$ band is expected to be dominated by intraband scattering\cite{Mazin2002}. If the kink position below T$_c$ is at  energy of $\Omega_{2g}+\Delta_{\sigma}$ \cite{Sandvik2004}, the energy shift should be $\Delta_{\sigma}$$\sim$7 meV.  This number should be even larger once the hardening of $E_{2g}$ mode reported by Raman measurements is taken into consideration \cite{Mialitsin2007}. Further work is needed to understand why the kink energy shift below T$_c$ is much reduced in MgB$_2$. 

The partial isotope coefficient $\alpha_M$ is defined as $\alpha_M=-\frac{dlnT_c}{dlnM}$, where $M$ is atomic mass (see, for example, Ref. \cite{Budko2001}). In standard BCS theory, $\alpha_M$ is equal to 1/2 \cite{Bardeen1957}, while the measured partial isotope coefficient of Boron in MgB$_2$ is substantially reduced to 0.26$\sim$0.31 \cite{Budko2001,Hinks2001}. The deviation of $\alpha_B$ from standard 1/2 can be understood by considering Coulomb repulsion pseudopotential ($\mu^*$) or strong phonon-phonon interaction (phonon anharmonicity) \cite{McMillan1968,Hinks2001}. Early theoretical \cite{Yildirim2001} and experimental \cite{Goncharov2001} results support the notion that large phonon anharmonicity exists in MgB$_2$, but were subsequently questioned by later reports \cite{Calandra2007,Simonelli2009}. In the harmonic limit, the phonon vibration frequency is directly proportional to the square root of atomic mass. We can use the kink energies measured by ARPES, which represent the energy of E$_{2g}$ phonon in the two isotopic MgB$_2$ samples, to check the harmonicity of this mode. The ratio of the measured the kink energy ($\frac{\omega_{B10}}{\omega_{B11}}=\frac{63}{66}=0.95 \pm0.03$) is very similar to the ratio of the square root of B atomic mass for the two isotopes ($\sqrt{\frac{M_{B10}}{M_{B11}}}=\sqrt{\frac{10}{11}}=0.95$).  These results suggest that the E$_{2g}$ phonon in  MgB$_2$ is likely in the harmonic limit.

In conclusion, we measured the isotope effect on the kink structure of MgB$_2$ superconductor. A shift of 3.5 meV is revealed between kink structure energy for B$^{10}$ and B$^{11}$ samples. 
Although the kink position shows a slight change below T$_c$ in both samples, the isotope effect on kink structure is not affected by superconducting transition. Our results provide  the first insights into the effects of the isotope substitution on electron-phonon interaction in a prototypical, conventional superconductor and will serve as baseline for understanding phonon effects in yet to be discovered novel superconductors.

This work was supported by the U.S. Department of Energy, Office of Science, Basic Energy Sciences, Materials Science and Engineering Division. Ames Laboratory is operated for the U.S. Department of Energy by Iowa State University under contract No. DE-AC02-07CH11358.

\bibliography{MgB2_iso}

\end{document}